\author{Aleksandra Andic \\
                Big Bear Solar Observatory\\
        40386 North Shore Lane\\
        Big Bear City, 92314, \underline{USA}
            \and
        J. Chae, K. Ahn\\
        Astronomy Program\\
        Department of Physics and Astronomy\\
        Seoul National University, Seoul 151-741, \underline{Korea}
        	\and
        P.R.Goode,  W. Cao, V. Yurchyshyn, V.Abramenko \\
                Big Bear Solar Observatory\\
        40386 North Shore Lane\\
        Big Bear City, 92314, \underline{USA}
}
\title{Response of Granulation to Small Scale Bright Features in the Quiet Sun}

\documentclass[12pt]{article}
 \usepackage{graphicx,multicol}

\begin{document}
\maketitle

\small{.\\
Submitted to ApJ: 20 August 2010\\
Accepted in ApJ: 7 February 2011}

\begin{abstract}
We detected $2.8$ bright points (BPs) per Mm$^2$ in the Quiet Sun (QS) with the New Solar Telescope (NST) at Big Bear Solar Observatory; using the TiO $705.68$ nm spectral line, at an angular resolution $~ 0.1''$ to obtain $30$ min data sequence. Some BPs formed knots that were stable in time and influenced the properties of the granulation pattern around them. The observed granulation pattern within $~ 3''$ of knots presents smaller granules than those observed in a normal granulation pattern; i.e., around the knots a suppressed convection is detected. 
Observed BPs covered $~ 5$\% of the solar surface and were not homogeneously distributed. BPs had an average size of $0.22''$, they were detectable for $4.28$ min in average, and had an averaged contrast of $0.1$\% in the deep red TiO spectral line.
\end{abstract}


\vskip7mm

\begin{multicols}{2}
{
\section{Introduction}
Observations of the solar photosphere performed in the G-band have revealed a plethora of tiny bright features, usually concentrated in active regions or bordering the supergranules in the QS (S\'{a}nchez Almeida et al. 2001, Berger et al. 2004, Rimmele 2004, S\'{a}nchez Almeida et al. 2004, Beck et al. 2007, Utz et al. 2009, Viticchi\'{e} et al. 2009, S\'{a}nchez Almeida et al. 2010, Viticchi\'{e} et al. 2010). Near disk center, they appear as "Bright Points" (BPs), roundish or elongated bright features located in the intergranular dark lanes (Dunn and Zirker 1973, Mehltretter 1974, Spruit 1979, Title et al. 1987, S\'{a}nchez Almeida et al. 2004). G-Band BPs are known to be associated with strong magnetic flux concentrations of about $1.5 \times 10^3$ G, hence, they can be used as a proxy to track small scale kG fields (Berger et al. 1995, 1998, van Ballegooijen et al. 1998, De Pontieu 2002, Beck et al. 2007,Viticchi\'{e} et al. 2009, Viticchi\'{e} et al. 2010). Thus, investigation of the properties of BPs can shed new light on the abundance of kG fields in the quiet Sun.\par 

Simulations of magnetohydrodynamics successfully reproduced the observable effects of magneto-convection on the solar surface. Models predicted formation of small-scale concentrations of the magnetic flux caused by the interaction of magnetic fields and convective motion (Spruit et al. 1990, Solanki, 1993, Shelyag et al., 2004, 2007). MURaM simulation code (V\"{o}gler et al. 2005) predict appearance of an increase in temperature in regions of down-flow (Shelyag et al. 2004 - Fig 3.) in the QS. A 3D LTE model (Hayek et al. 2010) provides a valid representation of the solar photosphere when scattering is not included in modeling of radiative transfer. Danilovi\'{c} et al.(2010) demonstrated quantitative agreement of synthetic observations made with MURaM code and real observations, indicating the validity of the simulations and the interpretations of the spectro-polarimetric observations.\par 
Additionally, Morinaga et al. (2008) observed that magnetic flux concentration may suppress convection in the active region; the strength of the magnetic flux is less relevant to the suppression of convection than the concentration of magnetic flux tubes. 
 The BPs are the smallest flux tubes that can be investigated because their sizes are often comparable to the resolution of the instrument used. The size of the BPs varies from $0.17''$ (Berger et al. 2004) to $0.8''$ (Berger et al. 1995), while Beck et al. (2007) found that the effective diameter of BPs is $0.2''$, concluding that very few of the BPs are larger than $0.4''$.  Utz et al. (2009) acquired two different results for the average diameter of BPs; $218$ km ($~ 0.3''$) when diffraction element was sampled at $0.108$ arcsec/pixel and $166$ km ($~ 0.22''$) when sampled at $0.054$ arcsec/pixel. The authors emphasize the influence of resolution on the apparent size of BPs.  Mehltretter (1974) found that BPs have typical dimensions of $100$ to $200$ km (from $0.13''$ to $0.27''$).\par 
Several methods for automatic detection of BPs have been developed (Bowelet and Wiehr 2001, 2007; Crockett et al. 2009; Utz et al. 2009). S\'{a}nchez Almeida et al. (2004) used a manual method to measure $0.3$ BPs per Mm$^2$ in the interior of the supergranulation cell, implying that BPs are ubiquitous in the photosphere. S\'{a}nchez Almeida et al. (2010) measured $0.97$ BPs per Mm$^2$ using the same method but better resolution. The number BPs detected is highly dependent on the resolution of the instrument and method used, consequently newer instruments and methods yield more BPs.\par 
 In this paper, we present evidence of a mutual interaction between small scale bright features and granulation. We report the observational properties of BPs (i.e., size, velocity and contrast) and compare them with previous studies.

\section{Data and analysis methods}\label{data}

Observations were performed with the New Solar Telescope (NST) at Big Bear Solar Observatory (Goode et al.  2010) on 29 July 2009. Photometry data of QS at the disk center were obtained using an optical setup containing a TiO broadband filter and PCO.2000 camera (Cao et al. 2010) in the Nasmyth focus.\par 
Data consists of a sequence of 120 bursts of 100 images. Each frame has an $10$ ms exposure time with a temporal cadence of $15$ s between bursts. Images have a sampling of $0.037''$ per pixel, which oversampled the diffraction element in the TiO spectral line. The field of view (FOV) encompassed $70'' \times 70''$ at the center of the solar disk. Image reconstruction and alignment of frames in the time series reduced the FOV to $54'' \times 54''$. Despite excellent seeing conditions, we were not able to reach the diffraction limit of $0.1''$ for TiO $705.68$ nm spectral line, because NST is still in commissioning phase.\par 
Images were speckle reconstructed based on the speckle masking method (von der L\"{u}he 1993). For this purpose we used the Kiepenheuer- Institut Speckle Interferometry Package (KISIP; W\"{o}ger et al. 2008)\footnote{The software package offers a GUI interface to choose reconstruction method and set parameters of reconstruction. We used the Triple Correlation Method with parameter settings: subfield = $5$, Phase rec. limit=$100$\%, Max. Iteration=$30$, SNR threshold=$90$, Weighing exponent=$1.3$ to complete image reconstruction. Wiener Noise Filter and Adaptive low pass filters were used.}.\par 
After speckle reconstruction, images were aligned using a Fourier routine. This routine uses cross-correlation techniques and squared mean deviations to provide sub-pixel alignment accuracy. However, we did not implement sub-pixel image shifting to avoid substantial interpolation errors that sometimes accompany this technique.\par 
We focused on the properties of small scale bright structures, e.g., their contrast, dimension, and lifetime.\par 
The term contrast is used as a measure of relative brightness as compared to the intensity of the QS, and is defined as:

\begin{equation}
C= \frac{I_m-I_q}{I_q},
\label{rms}
\end{equation}

\noindent where $I_m$ is the mean intensity of all pixels from the targeted image area and $I_q$ the mean intensity of the truly QS in our FOV. FOV contained a plethora of BPs so we defined QS in following way:  As a QS we chose subareas that did not contain any of the resolvable small structures (Fig.\ref{primjer}A). Intensity from these areas was used to obtain the QS intensity. This provided us with the intensity of QS that had no resolvable contribution of the small structures. Fig. \ref{primjer}B shows an example of one of the studied objects.}

\end{multicols}

\begin{figure}[h]

\includegraphics[scale=0.4]{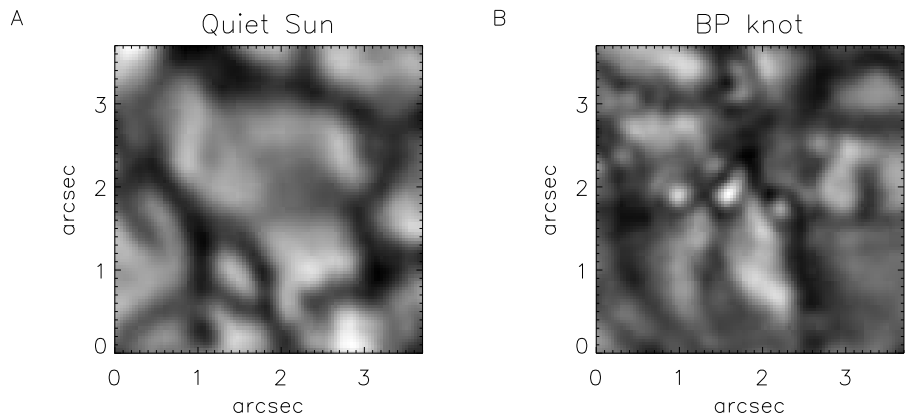}
\caption{Panel A presents an example of the quiet sun as we defined it for this work. Panel B presents one of the stable locations of the BPs in our field of view, as the contrast to the QS. }
\label{primjer}
\end{figure}

\begin{figure}[h]
\includegraphics[scale=0.55]{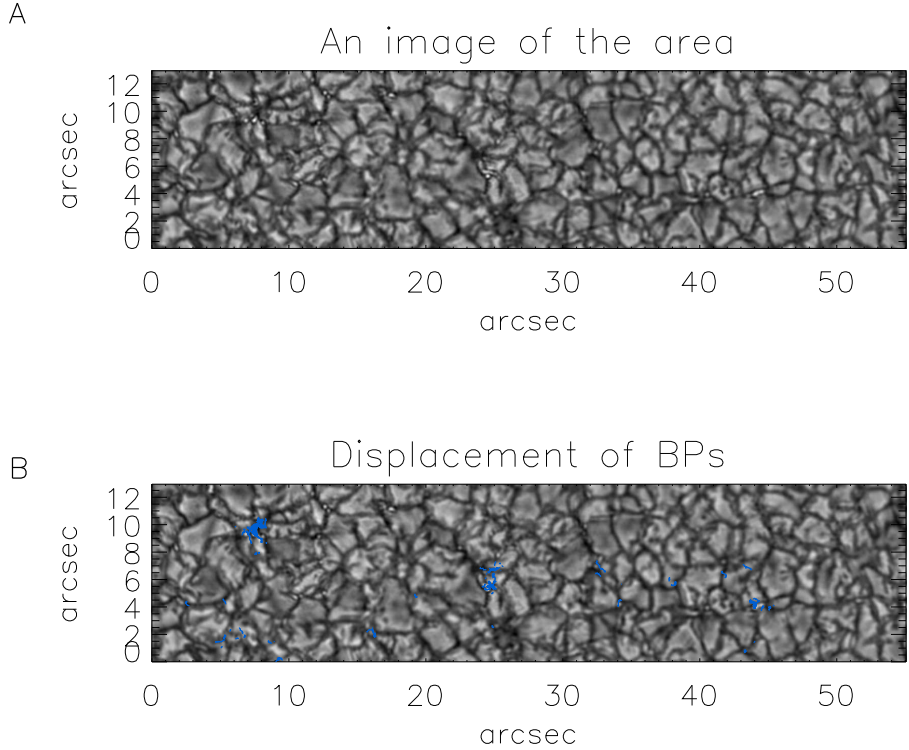}
\caption{An individual image of the selected region for analysis (panel A) and displacement of the analyzed BPs during their individual lifetime (panel B). Over granulation image the traces of BPs movement in time were plotted (blue solid line).  Persistent conglomerations (i.e., knots) of BPs are visible at coordinates ($8''$,$9''$), ($25''$,$6''$), and ($45''$,$4''$).}
\label{intensities}
\end{figure}

\begin{multicols}{2}
{
We measured diameter of BPs when they were the brightest to achieve greater accuracy and took intensity profiles using full-width at half maximum (FWHM). For aspheric BPs we measured FWHM from light profile on the longest axis. \par 
We calculated feature velocities using the nonlinear affine velocity estimator (NAVE) method (Chae and Sakurai, 2008), which applies an affine velocity profile instead of a uniform velocity profile commonly used in the local correlation tracking (LCT) method.\par 
Area coverage and number density were obtained using the following methods. We first isolated intergranular space using part of the procedure developed by Crockett et al. (2009). While Crocketts procedure is optimal for some telescopes, the higher resolution of the NST required us to use a modified version  method of S\'{a}nchez Almeida et al. (2004). Instead of playing series back and forth to single out the BPs we used NAVE method to track the plasma flow of individual BPs that could be tracked for at least 2 frames and were at least $0.13''$.\par  
We used the wavelet technique to obtain the feature size power spectrum. The wavelet technique was applied following the constraints described in Andic (2010), however, we increased the confidence level to $99$\% to reduce the influence of noise induced irregularly across the FOV by the PCO.2000 camera, most prominently in the upper left half of FOV.\par 
Wavelet analysis is a very efficient filter;it separates different period oscillatory signals from processed signal (Torrence and Compo, 1998) forming a set of sub-ranges of oscillations across the main period range. For our spatial analysis the main range is  $0.076''$ - $11''$, and analysis formed 60 sub-ranges. Only periodic signals that match set conditions are kept. Wavelet automatic analysis calculates confidence levels for every detected oscillatory signal. False detections are reduced to the minimum when confidence levels are set to $99$\%. The finite nature of the signal will cause false detection near the edge of signals. These false detections can be removed by ignoring all oscillatory detections near the edge of a signal, an effective technique for shorter period oscillatory signal, but not for the longer periods. Prolonging a signal is the only viable option for the longer periods.  We chose the size of sub-areas to $300 \times 300$ pixels  as optimal for our study, because we can notice the change in granulation shape and false detections are prominent only for structures of $\geq 3''$ . \par 
We selected a segment of the FOV (Fig. \ref{intensities}A) and analyzed $112$ BPs over time.  A segment was selected to avoid camera-induced noise, most prominent in the upper left portion of the FOV. Withing this segment all BPs who appeared during the time sequence and matched set conditions (size $\geq 0.13''$ and time of the detection $\geq 30$ sec) were analyzed. Camera noise resulted in numerous artifacts of $~ 0.1''$, which would undermined the reliability of our analysis. 
}
\end{multicols}

\begin{multicols}{2}
{
\section{Results}

\subsection{Spatial distribution of BPs}

The spatial distribution of BPs across our FOV was analyzed following methodology presented in S\'{a}nchez Almeida et al. (2004, 2010) in a frame that had the highest difference between intensity of the bright and dark features.  Because we counted more BPs than reported in previous research (S\'{a}nchez Almeida et al. 2010), we repeated analysis on $7$ more frames to increase reliability of our results. We noted on average $743.62$ BPs per frame, yielding $2.8$ BP per Mm$^2$.\par 
}
\end{multicols}

\begin{figure}[h]
\centering
\includegraphics[scale=0.5]{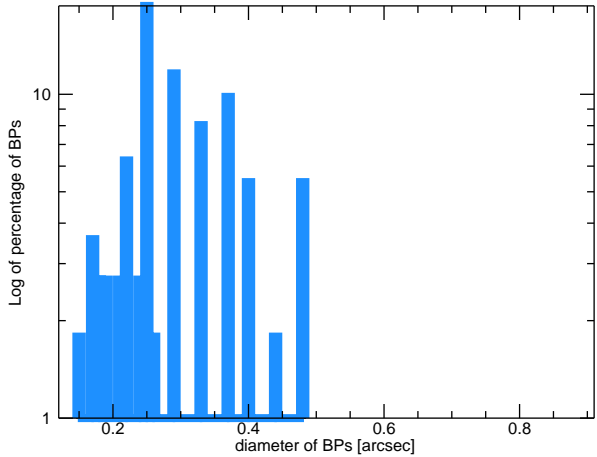}
\caption{Size distribution of analyzed BPs. Y axis presents percentage of BPs in logarithmic scale. The largest percentage of BPs has $0.22''$ and that value is also close to the average value of diameter of all measured BPs.}
\label{velicina}
\end{figure}

\begin{figure}

\includegraphics[scale=1]{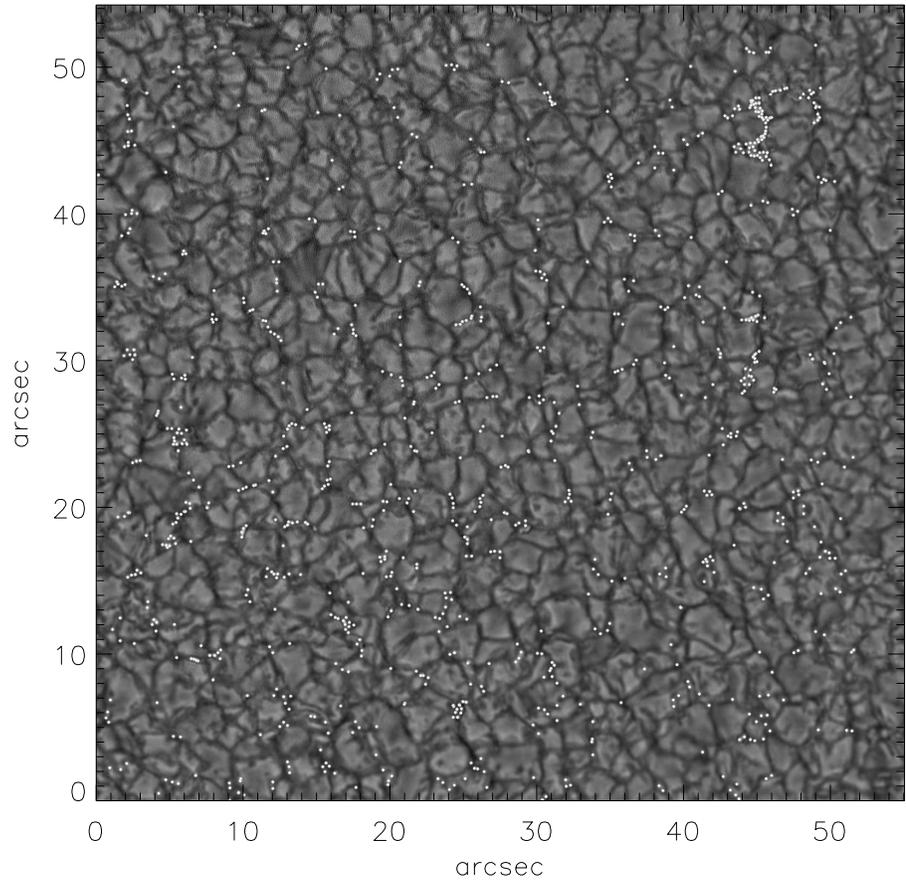}
\caption{Analyzed frame from dataset. We increased visibility of BPs by multiplying their intensity by a normalized Gaussian function with half-width $0.14''$. }
\label{fov}
\end{figure}

\begin{multicols}{2}
{
Percent coverage of a FOV depends on BP size (Fig. \ref{velicina}). If we assume $0.22''$ as the average size, BPs cover $~ 5.38$\% of the FOV. This spatial distribution is the largest observed so far (S\'{a}nchez Almeida et al. 2010). S\'{a}nchez Almeida et al. 2004 stated that BPs are ubiquitous in the photosphere of the QS. We cannot confirm this, since we did found several patches of the $3.7'' \times 3.7''$ QS where we could not resolve any small features (Fig. \ref{fov}). 
}
\end{multicols}

\begin{figure}[h]
\centering
\includegraphics[scale=0.4]{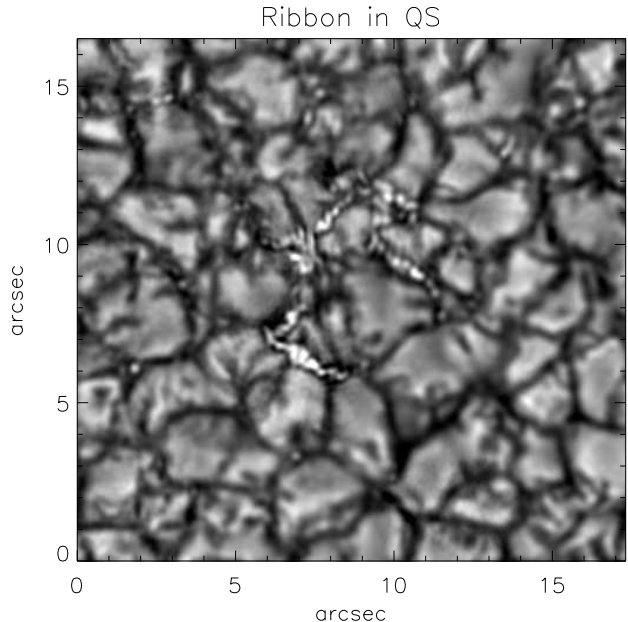}
\caption{A zoomed out right top corned of FOV. This structure is called a ribbon and it is considered to be conglomerate of the BP in areas with strong magnetic flux. With NST resolution we can resolve individual BPs and clusters of BPs that compose this ribbon.}
\label{traka}
\end{figure}

\begin{multicols}{2}
{

\subsection{BPs concentrations}

A majority of BPs were part of persistent knots that remained stationary during our observations. We define "knots" as locations where persistent BP's are observed for the duration of the time series (Fig. \ref{intensities}B).\par 
The change in size distribution of granules could point to an influence of a constant presence of a BPs concentrations.  We choose 4 distinct areas of the same size ($300 \times 300$ pixels) for wavelet analysis; the QS (Fig. \ref{primjer}A) and  area with ribbon as reference areas (Fig. \ref{traka}), and  two areas with knots.\par 
We noted a structure stable during the time series (Fig. \ref{traka}), with the shape similar to ribbons previously noted in moth areas (Berger et al. 2004). Our observations were preformed during the solar minimum, and there were no active regions present. Also, during our observations, SOHO MDI was not providing data. For several days prior to and after our observational run, hoverer,  in the same region MDI magnetograms showed only "salt and pepper" flux distribution typical for the QS regions. One observed ribbon and lack of magnetic information makes our dataset inappropriate for analyzing in detail this structure. We used this region as the control region, since it contained a plethora of BPs.

}
\end{multicols}

\begin{figure}[h]
\centering
\includegraphics[scale=0.5]{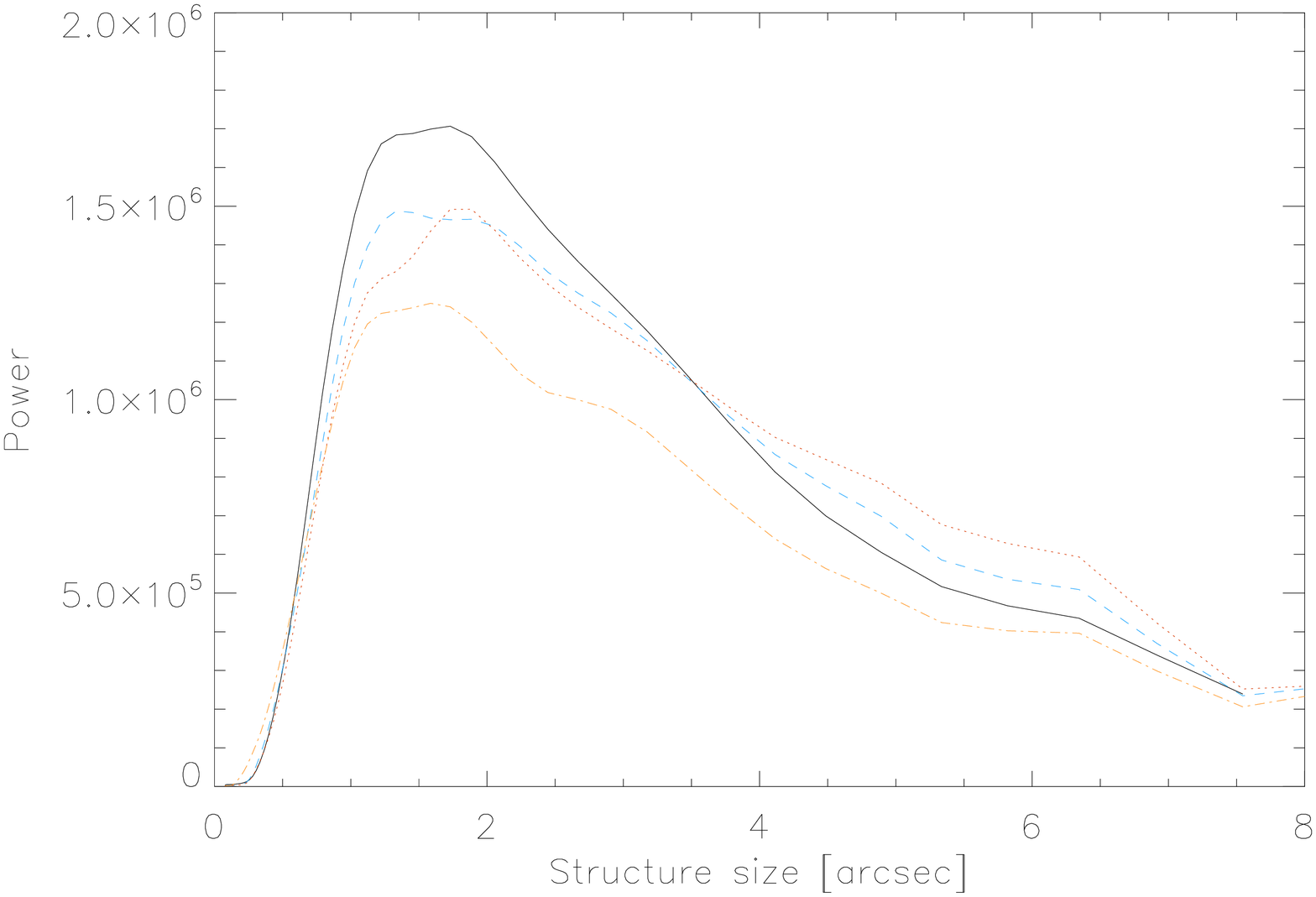}
\caption{Result of wavelet analysis of sub-images. Black solid line represents distribution of different sized structures in the granular subfield of QS.  Blue dashed line represents same for sub-field that contains one of the analyzed knots, while dotted red line other knot. Orange dot-dashed line represents distribution of structures in subfield containing the ribbon.}

\label{velici}
\end{figure}

\begin{multicols}{2}
{

QS area (Fig. \ref{primjer}A) was expanded to $300 \times 300$ pixels to minimize an edge effects induced by wavelet analysis. This expansion included sample of BPs in the QS. Nevertheless, even with additional BPs this reference area remained the quietest.\par 
Two remaining areas were $300 \times 300$ pixels areas, each containing  knots  from coordinates: ($8''$,$9''$) and ($25''$,$6''$), respectively.\par 
We analyzed the size of structures in each sub-area using wavelet analysis, yielding distribution of sizes (Fig. \ref{velici}). Analysis was done separately for each frame in the time series and results are integrated over the time. 

}
\end{multicols}

\begin{figure}[h]
\centering
\includegraphics[scale=0.4]{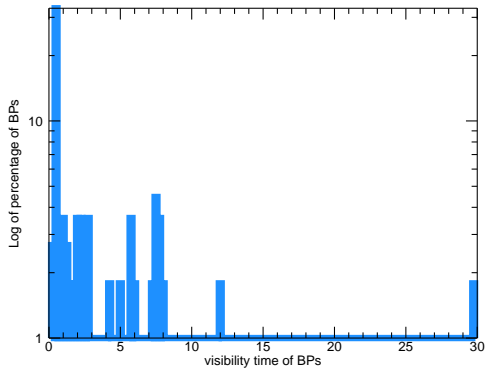}
\caption{Distribution of visibility time of BPs within the dataset. Y axis presents the percentage of the analyzed BPs in logarithmic scale, x axis the time of the visibility. The visibility time of BPs do not match the true lifetime of the structures.}
\label{zivot}
\end{figure}

\begin{multicols}{2}
{
The modal granular size was QS is $1.7''$, with the smallest observed structure as $~ 0.2''$ (Fig. \ref{velici}, solid black line). The sub-area containing a knot had $11.32$\% fewer structures in $0.8''$ - $2.7''$ than QS and the mode was $1.3''$ (Fig. \ref{velici}, blue dashed line). The sub-area containing the second knot (Fig. \ref{velici}, red dotted line) contained $15.48$\% fewer structures in  $0.8''$ - $2.7''$ than QS and the mode was $1.8''$.\par 
An area with a ribbon has $1. 6''$ granule as the mode with an abundance of structures $\leq 0.8''$ (Fig. \ref{velici}, orange dot-dashed curve). This area contained $26.46$\% fewer structures in $0.8''$ - $2.7''$ than QS. The number of BPs per Mm$^2$ in this area is $8.59$. The size distribution curve for the area shows a noticeable increase in power for structures $ ~ 3''$. This increase is in part caused by the dimensions and shape of the ribbon in addition to edge effects. \par 
Influence of conglomerations of BPs is most prominent in area with a ribbon; which had $30.12$\% more structures in $0.12''$ - $0.56''$ than QS, while areas with knots had $3.54$\% more structures in $0.12''$ - $0.56''$ than QS.\par 
The dataset limits the possibility of connecting locations of knots to a network or intranetwork, because the position of network in the FOV cannot be determined precisely (Fig. \ref{fov}). In areas with knots, there were $3.24$ BPs per Mm$^2$ (Fig. \ref{velici}, blue dashed line) and $3.29$ BPs per Mm$^2$ (Fig. \ref{velici}, red dotted line) making it $~ 16$\% greater than average for the whole FOV. 

\subsection{BP statistic} 

A majority of analyzed BPs were in  $0.13''$ - $0.48''$ (Fig. \ref{velicina}). Larger BPs are rare; only $6$\% of all analyzed BPs were larger than $0.48''$. The mode was $0.22''$ comprising $19$\% BPs, while  the average diameter of BP was $0.22''$.\par  
The lifetime of BP is influenced by abilities of instruments. Because BPs sizes are strongly affected by instrumental resolution, measured lifetime is actually the product of BP visibility to instruments rather than the period between their physical creation and disappearance.
BPs average lifetime is $4.28$ min, while median lifetime was $~ 2$ min, in a range of our imposed minimum of $30$ s to the longest lifetime of $29.75$ min, while the modal lifetime is $0.5$ min (Fig. \ref{zivot}). The large percentage of BPs with lifetime of $30$ s is likely a consequence of an imposed minimum.  We also detected $2$\% of BPs that were traceable trough their plasma flow for $~ 30$ min. All long-lived BPs were located in the knots. 

}
\end{multicols}

\begin{figure}[h]
\centering
\includegraphics[scale=0.5]{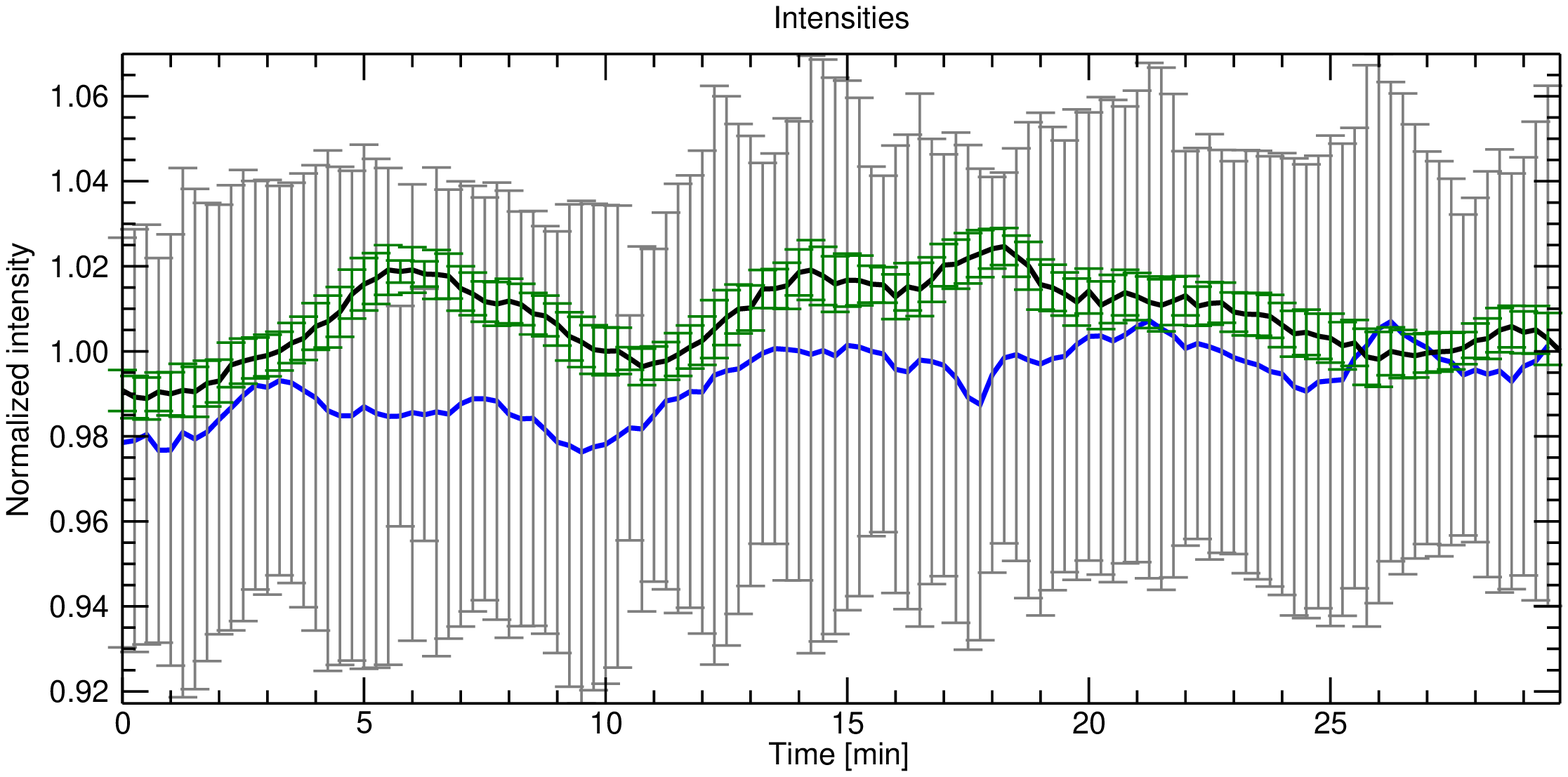}
\caption{Intensities of QS and BPs, normalized on a mean intensity of flat field frame. Black line with green error bars represent intensity of QS with corresponding errors. Blue line with gray error bars represent intensity of BPs.  The intensity of BPs was obtained by measuring intensity changes of each BP and then averaging them. Intensity error is calculated using IDL procedure for the subsequent pixels. Error is equal to values where standard deviation of convolved area was smaller than standard deviation of same area convolved with a transposed kernel. These intensity profiles indicate that at TiO continuum formation level, intensities of BPs and granulation are similar.}
\label{intenzitet}
\end{figure}

\begin{multicols}{2}
{
Intensity of each frame was normalized to the mean intensity of a flat field frame. We measured intensity of QS (Fig. \ref{intenzitet}, black line) and compared it with the intensity of BPs. (Fig.\ref{intenzitet}, blue line).  Smaller areas of BPs caused larger errors during the intensity measurements. The QS intensity curve with its errors lie inside the error range for the intensity of the BPs. Thus, we can state that brightness of the BPs and QS are equal at the formation level for the continuum of the spectral line TiO $705.68$ nm.

}
\end{multicols}

\begin{multicols}{2}
{
Contrast averaged $0.1$\% for BPs, however, intensity error for the data-set was $0.07$\%, hence we can state that BP contrast in the continuum of the spectral line TiO  $705.68$ nm is equal to the QS contrast.

\section{Discussion}

\subsection{Spatial distribution of BPs}

We used a broadband filter centered at TiO $705.68$ nm spectral line, previously used only for an umbral investigations because it is very weak above granulation. Although use of TiO spectral line and broadband filter provide images consisting mainly of line continuum, our results can provide a reliable estimate of the distribution of small scale structures though not necessarily as a proxy for intense magnetic concentrations (Shelyag et al. 2004, Beck et al. 2007). No detailed analysis of the connection between magnetic properties of small scale structures in TiO line has been done to our knowledge.\par 
A BP $\geq 0.13''$ should appear in at least $2$ sequential frames to be considered for our analysis. NST has a clear aperture of $1.6$m, achieving resolution close to $0.1''$ in the TiO $705.68$ nm spectral line. However, the camera used and the speckle reconstruction method induced artefacts of a size approaching the diffraction limit, hence we ignored all structures smaller than $0.13''$. We observed $2.8$ BPs per Mm$^2$, $\sim 3$ times larger than the previous estimate for the G-band spectral line (S\'{a}nchez Almeida et al. 2010). Since we used very similar method as one used by S\'{a}nchez Almeida et al. (2004, 2010), we can speculate that differences in our results were caused by the different spectral line used. 
 
\subsection{BP concentrations}

BPs showed a tendency to group at several locations in the FOV, which we call "knots". We analyzed $2$ persistent BPs knots at the center of the solar disk (Fig. \ref{intensities}). Knots seem to influence granulation in their immediate vicinity by causing appearance of smaller granules (Fig. \ref{velici}). Power spectrum of structure sizes in $4$ analyzed sub-areas showed a drop in power for structures in  $0.8''$ - $2.7''$ in sub-area that with high concentrations of BPs.\par 
In subfields around knots the number density of BPs was $16$\% larger than average for the whole FOV. In both areas with knots, BPs cover $0.5$\% and $0.6$\% more than average, indicating that mere spatial distribution of BPs cannot be the cause of the different size distribution (Fig. \ref{velici}) in those areas; rather their clustering around the knot location caused drop in power for structures in $0.8''$ - $2.7''$. On the other hand, the area with a ribbon has $8.59$ BPs per Mm$^2$ and BPs cover $12.8$\% of sub-area ($11.1'' \times 11.1''$). In this case the shape of the size distribution curve (orange dot-dashed line, Fig.\ref{velici}) is influenced by BP numbers too and not only by flux concentrations.  
 We can conclude that size of granules appears to be affected by the presence of a concentration of BPs in QS.\par 

Suppression of convection by magnetic fields was known for decades (Parker, 1978). Morinaga et al. (2008) established that convection is suppressed by high concentrations of magnetic flux tubes, not by strength of magnetic fields. Arguments suggest that some kG flux tubes are not bright in the G-band (S\'{a}nchez Almeida et al. 2001, V\"{o}gler et al. 2005, Beck et al. 2007). This agrees with Ishikawa et al. (2007), who found areas of the high magnetic field without BPs.  These arguments may indicate that higher resolutions are capable of resolving more kG flux tubes as BPs.\par 
If we assume that BPs in TiO continuum are proxies for intense magnetic concentrations, we can speculate that our result is also based on flux tube concentration; i.e. the noted drop in the granulation size is probably caused by the concentration of the flux tubes in the knot.\par 
This suppressing is localized to the vicinity of a BP knot. This effect is easily noticeable and measured in active regions and network patches, since BPs are found over many arcseconds. It is more complicated to detect this effect in the QS since the cluster dimensions are usually comparable to the resolution of the instrument. With our increased resolution we were able to measure those effects. \par 

\subsection{BP statistic}

Crockett et al. (2010) stated that magnetic BPs cannot be generated in large diameter magnetic flux tubes, and Beck et al. (2007) concluded that very few BPs are larger than $0.4''$.  This is consistent with our observations that only $6$\% BPs were larger than $0.5''$. With the resolution achieved in this dataset, we cannot see the real shape of an average BP, since the smallest resolved BPs are equal to our set limit. \par 

Viticchi\'{e} et al. (2010) put forward the upper limit of $\sim 0.1''$ as the dimension over which the elementary G-band bright features are formed. With our observations we could neither confirm nor dispute this. The average size of BPs in our dataset is two times larger, and the smallest BPs we observed was larger than limit put forward by Viticchi\'{e} et al. (2010). \par 

The size of BPs in this dataset agrees with the result of Utz et al. (2009) and Beck et al. (2007). Utz et al. (2009) obtained different mean values for different sampling sizes (with sampling of $0.054$ arcsec/pixel authors measured $166 \pm 31$ km). This agrees with our result of BPs diameter of $0.22''$ measured with sampling of $0.037$ arcsec/pixel. Beck et al. (2007) observed BPs with an average diameter comparable with our result; differences in numbers were most likely caused by different sampling sizes. \par 

Rimmele (2004) observed $30$\%-$50$\% larger apparent size of BPs in the G-band than in the continuum. Although we analyzed BPs near continuum, we cannot confirm this result. We found that mean diameter of BPs is comparable with the size of BPs in the G-band reported by Rimmele (2004). \par 

Berger et al. (2004) observed larger, amorphous ribbons in an active region plage near disk center. We observed a similar structure in the QS. Unfortunately, at the day of our observations there was no magnetogram obtained by MDI, SOHO. The magnetograms from the previous and following days show the magnetic flux distribution typical for QS in the area where the ribbon was located. The difference between our structure (Fig. \ref{traka}) and the one analyzed by Berger et al. (2004) is that we could see our ribbon resolved to the individual BPs and clusters and ours was located in the QS. However, from analysis done with this dataset we cannot offer any explanation of this structure. Future analysis may benefit from inclusion of magnetic and Doppler-velocity information. \par 
The lifetime of BPs analysed is strongly influenced by the resolution of the instrument and the duration of the dataset. We had to impose lower temporal ($30$ s), and spatial ($0.13''$) limits on the observed structures, limiting what we were able to see.  The duration between physical appearance and disappearance of BPs will be possible to establish only when spatial resolution of instrument  is high enough to resolve an individual flux tube and follow it from formation to disappearance. In our dataset, the moment when BPs appear or disappear does not correspond to BPs actual formation or disappearance of BPs, but is instead a measure of the visibility of BPs within the limits of our instruments. 

\subsection{Irradiance}

Variation in total solar irradiance is caused by magnetic fields at the solar surface (Chapman et al. 1996, Lean et al. 1998, Filgge et al. 1998). Krivova et al. (2003) argued from their model that solar surface magnetism is responsible for solar irradiance changes. This argument can be tested by measuring contrast of BPs, pronounced in molecular lines along the blue side of the visible spectrum and diminishing as one goes toward the red side of the spectrum. At near-infrared wavelengths contrast reaches negative values (Tritschler and Uitenbroek, 2006). According to Planck's law, for a deep red spectral line, a mean contrast of BPs at the solar disk center should be very close in value to the contrast of the QS.  Tritschler and Uitenbroek (2006) presented contrast measurements for several spectral lines that followed the prediction from Planck's law. \par 
The broadband TiO filter allowed us to access a $\tau_{500}=1$ level, where we can expect hotter granules than in G-band. Hence, contrast of BPs observed with this filter should be closer to the contrast of the granules (Fig. \ref{intenzitet}).\par 
The frequency range of the broadband TiO filter falls into the deep red part of the visible spectrum. Tritschler and Uitenbroek (2006) analysed spectral lines that fall around the deep red part of the visible spectrum. Considering the scale Tritschler and Uitenbroek (2006) established, our finding that the contrast of BPs is equal to the contrast of the QS, is in agreement with their results. Hence, there is a possibility that even BPs in the quiet Sun make a contribution to the global solar irradiance.

\section{Summary}

Small concentrations of BPs seem to cause a reaction of convection in the quiet Sun, slightly suppressing convection in their immediate vicinity. \par 
The NST at BBSO resolved a plethora of BP-like structures in the QS, around $2.8$ BPs per Mm$^2$ with TiO spectral line. Those structures tend to form persistent knots, which in turn forced the shape of granulation pattern to adjust themselves around the knot location over the duration of the observational run. Size distribution of granules in the immediate vicinity of knots points to a slightly suppressed convection due to the increased concentration of BPs. \par 
On average, analyzed BPs are $0.22''$ in size and detectable for $4.28$ min. Their averaged contrast is the same as the contrast of the QS. \par 

\emph{Thanks are due to the anonymous referee whose comments helped improving this work. AA thanks J.T. Villepique  for help with English.
We gratefully acknowledge support of NSF (ATM-0745744 and ATM-0847126), NASA (NNX08BA22G) and AFOSR (FA9550-09-1-0655). }

\bibliographystyle{abbrv}
\bibliography{simple}

}
\end{multicols}

\end{document}